# A scale space based algorithm for automated segmentation of single shot tagged MRI of shearing deformation

André M.J. Sprengers MSc[1], Matthan W.A. Caan PhD[1], Kevin M. Moerman MSc[1], Aart J. Nederveen PhD[1], Rolf M.J.N. Lamerichs PhD[1,2], Jaap Stoker MD PhD[1]

[1]Department of Radiology, Academic Medical Center, University of Amsterdam, Amsterdam, the Netherlands,
[2]Philips Research, Eindhoven, the Netherlands

## Abstract

*Object* This study proposes a scale space based algorithm for automated segmentation of single-shot tagged images of modest SNR. Furthermore the algorithm was designed for analysis of discontinuous or shearing types of motion, i.e. segmentation of broken tag patterns.

*Materials and methods* The proposed algorithm utilizes non-linear scale space for automatic segmentation of single-shot tagged images. The algorithm's ability to automatically segment tagged shearing motion was evaluated in a numerical simulation and in vivo. A typical shearing deformation was simulated in a Shepp-Logan phantom allowing for quantitative evaluation of the algorithm's success rate as a function of both SNR and the amount of deformation. For a qualitative in vivo evaluation tagged images showing deformations in the calf muscles and eye movement in a healthy volunteer were acquired.

*Results* Both the numerical simulation and the in vivo tagged data demonstrated the algorithm's ability for automated segmentation of single-shot tagged MR provided that SNR of the images is above 10 and the amount of deformation does not exceed the tag spacing. The latter constraint can be met by adjusting the tag delay or the tag spacing.

*Conclusion* The scale space based algorithm for automatic segmentation of single-shot tagged MR enables the application of tagged MR to complex (shearing) deformation and the processing of datasets with relatively low SNR.

# 1 Introduction

The use of tagging in Magnetic Resonance Imaging (MRI) was first proposed for cardiac imaging [1, 2]. By saturating bands of zero signal in the image, tagging allows for imaging of cardiac wall motion and provides quantitative information on cardiac function. Since then, tagged MRI has been applied to various other fields of research and regions of the human body such as the eye [3, 4], tongue [5, 6] and skeletal muscles [7–9]. The main difference in applying tagged MRI to body regions other than the cardiac region is the lack of periodicity. However, tagged MRI can still be used by repeating the movement of interest [10] or—when repetition is not possible or not desirable—the tagged acquisition has to be single shot and real-time [9, 11, 12]. As a consequence, SNR improving techniques such as C-SPAMM [13, 14] or extending the scan duration are unavailable and a tradeoff between temporal and spatial resolution has to be made. The use of tagging outside the heart is further complicated by sliding or shearing edge deformations, which tend to occur when imaging motion of independently moving parts such as skeletal muscles (and skeletal muscle parts) or bowel motion. Sliding and shear deformations may cause tag lines to break up and separate at the interface between moving parts. Such complex deformations severely hinder automated segmentation of the tag lines. In addition, there is no option of obtaining the tag pattern in increasing phases of deformation, since the motion is non-periodic. Segmentation errors can be corrected manually, but this becomes problematic and time consuming when a large number of readouts are required, e.g. where physiological phenomena have a long time span, and can hamper the clinical implementation. In this study we propose an automated algorithm based on nonlinear scale space for tag segmentation of continuously tagged image series, i.e. a large number of single shot tagged dynamic volumes. The algorithm was evaluated for tag segmentation under shearing edge deformations in a numerical phantom and in vivo for active calf muscle deformations and eye movement in a healthy volunteer.

# 2 Materials and methods

We aimed to develop an algorithm capable of automatically segmenting continuously tagged image series for estimating discontinuous (shearing edge-like) and non-repetitive motion patterns in data with relatively low signal-to-noise ratios (SNR). The algorithm was tested for its ability to segment tags deformed by simulated shearing edge deformations in a numerical phantom and in vivo deforming calf muscle of a healthy volunteer.

## 2.1  Numerical phantom dataset

A tagged shearing edge deformation was simulated in a 2D Shepp-Logan phantom. Tag lines and Gaussian noise were added to the phantom image, after which the image was distorted causing a shearing edge. To methodically evaluate the algorithm's capability, the level of noise and shearing magnitude were both varied, covering the range expected for SNR and typical shearing in tagged MR acquisitions. Analogous to tagged MRI, a horizontal tagging pattern was generated with a sinusoidal modulation:

$$M(y) = |m + A \cdot \sin(2\pi \cdot \tfrac{y}{d})| \tag{1}$$

corresponding to a 1-1 tag pulse, *m* and *A* are the mean value and amplitude of the oscillation respectively, and *d* represents the spacing between the tag lines. For a straightforward simulation of shearing-like motion in human tissue, the phantom image was distorted with a calculated vector displacement map *D* whose origin coincides with the centre of the image. The displacement map *D* is composed in two steps. First, a shift was introduced to the origin of the map, the magnitude of which decays exponentially with the distance from the origin according to:

$$D(r) = h \cdot e^{\left(\frac{|r|^2}{2\alpha^2}\right)} \tag{2}$$

where *h* defines the direction and magnitude of the shift, $|r|^2$ is the distance of a given location *r* in the vector map with respect to the origin and $\alpha^2$ is the decay rate of the distortion. Second, a shearing effect is introduced into *D* by reversing the direction of all vectors on the right side of the map ($D_{x>0}(r) = -D_{x<0}(r)$). For the purpose of generating a basic shearing deformation, we define the vector *h* in the positive z direction (upwards), perpendicular to the tagging pattern with a maximum magnitude of one tag spacing (*d*). The resulting displacement map D for the shearing edge deformation is shown in Figure 1a. Before implementation of the tag pattern and applying the shift (or expansion) map, the resolution was increased by a factor of 10, i.e., each original pixel is replaced by 100 pixels which each have the same value. The shift map was generated with that same increased resolution. After implementation of the tag pattern, the distortion was applied in this resolution by shifting each pixel to new coordinates according to D. This produced a list of points with real x and y coordinates and a weight corresponding to the pixel intensity. This list was then used to construct a new distorted image with the original resolution. In order to do this, for each item in the list, the weight will be distributed over the four nearest pixels (where the distribution is proportional to the distances of the point to each of the pixels) after which the image was scaled back to the original resolution. This procedure guaranteed the smoothness of the map and conserved the total intensity of the image.

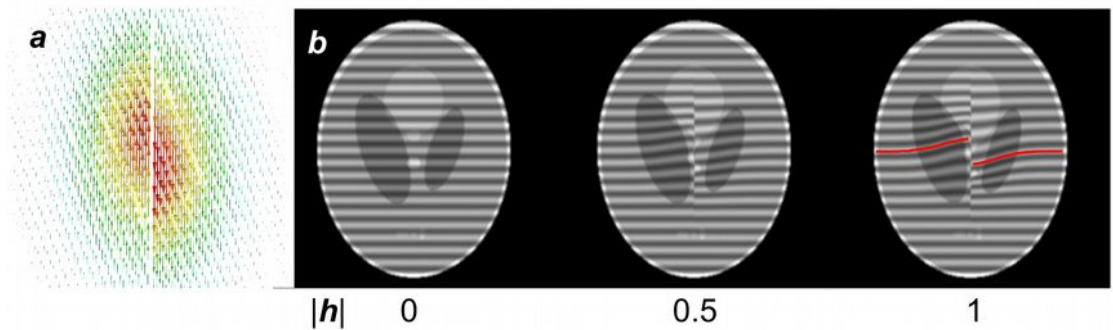

***Figure 1:*** *Simulation of a shearing deformation. **a** Shows a vector representation of the deformation map used to effect a shearing deformation in a Shepp-Logan phantom. **b** Shows an example of a deformed Shepp-Logan phantom with a central shift of |**h**|=0, 0.05 and 1 d (d = tag spacing). The red line at the right of **b** indicates the amount of tag separation in the centre of the image at |**h**| = 1.*

Figure 1b shows the tagged numerical phantom with a distortion of |**h**| = 0, 0.5 and 1*d*. For systematic evaluation of the segmentation algorithm, series of shearing edge deformation in the Shepp-Logan phantom were generated with SNR ranging from ∞ to 1.5 and shifts ***h*** with magnitudes ranging from zero to the maximum magnitude *d*. Note that because of the shearing effect, at full deformation a net separation of twice the tag spacing is observed in the origin of the image (see the right side of Figure 1b). For the purpose of error calculation, 10 noise realizations were generated for each shift and SNR level. The SNR level was defined as the ratio of the mean and standard deviation of the intensity over the complete set of images.

*2.2 A scale space based algorithm for tag line segmentation*

The scale space framework is a commonly used approach in image processing to robustly segment an object's noisy data [15, 16]. Images are blurred with Gaussian kernels of increasing size, generating a series of images with varying scale dimension (i.e. width of the Gaussian kernel). At high blurring scale, detailed features are suppressed and the remaining global structures can be segmented. By combining the information gained from each scale dimension, a more successful segmentation of the image can be achieved. We developed a scale space based algorithm suited especially for the typical features encountered in tagged image series. The algorithm operates in 2D. Per slice of a dynamic volume, a series of scale space representations of the original image at different levels is created. In the conventional, linear scale space framework an isotropic Gaussian kernel is adopted. For the purpose of segmentation of the deformed taglines, we impose an anisotropic kernel, oriented along the tag orientation (the *x*-direction). In this case the (non-linear) scale space may be formulated as:

$$L(x,y,\sigma_x) = g(x,y,\sigma_x) * I(x,y) \quad \text{with} \quad g(x,y,\sigma_x) = \frac{1}{2\pi} e^{\frac{x^2}{2\sigma_x} + \frac{y^2}{d}} \quad (3)$$

with $L(x,y,\sigma_x)$ the derived image $I(x,y)$ at scale dimension $\sigma_x$, $g(x,y,\sigma_x)$ the anisotropic blurring kernel and *d* the tag spacing. The scale dimension ranges from twice the width of the image to zero in steps of one voxel (meaning no blurring is applied) (see Figure 2a).

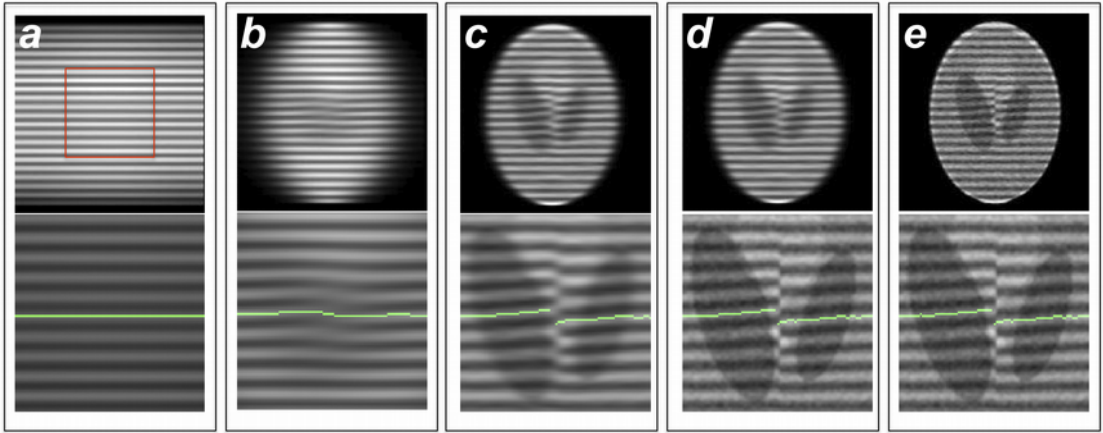

*Figure 2: Generation of the scale space dimension. In **a**, the scale space dimension is maximal, i.e., the full width at half maximum (FWHM) of the blurring kernel equals twice the width of the image. In the lower half of Figure2, enlargements are displayed to the region indicated with the red square in the upper half of **a**. Segmentation of the middle tag is indicated in green. In **b-e** the FWHM equals 50, 10 and 5 voxels. The scale dimension of the right frame equals one, meaning that no blurring kernel was applied.*

To automatically segment the tag pattern, first the tags are initialized by segmenting the highest dimension of scale space $L(x,y,\sigma_x)$ (see Figure 2a). Due to the large blurring extent, the tagging pattern will appear linear at this scale. Also the tagging pattern is dominant over any detailed structure or noise in the image. Thus, segmentation at this scale is performed by selecting the voxels with maximum intensity with respect to neighboring voxels perpendicular to the tag direction; no further assumptions about the tag spacing or the shape of the tag pattern is used to segment this lowest scale.

$$L_{seg}(x, y_i, \sigma_{x\_max}) = \arg\max(L_{seg}(x, y_j, \sigma_{x\_max})|j = \{i-1, i, i+1\}) \quad (4)$$

The image with the second highest blurring scale (Figure 2b) is segmented by selecting for each of the voxels found in the previous dimension the voxel with the highest intensity within a one-voxel vicinity along the direction perpendicular to the tag pattern initial orientation:

$$L_{seg}(x, y_i, \sigma_n) = \arg\max(L(L_{seg}(x, y_j, \sigma_{n+1})|j = \{i-1, i, i+1\}) \quad (5)$$

This step is repeated for all scales (Figure 2c–e), resulting in an automatic segmentation of the original non-blurred image. For each segmented tagged voxel a maximum intensity has to be present within a one-voxel vicinity for each scale dimension. Within this procedure, each point located at the maximum of the deformed tag pattern is segmented individually. No assumptions on the nature of the deformation are introduced, allowing for separation or shearing of tagged voxels to occur. The procedure is repeated for all slices and dynamic volumes. Because of the non-linear framework of the scale space algorithm, there is a chance of creating false or spurious edges along the dimensions of scale space [15, 16]. In this particular case this will occur when tags blend i.e. when the amount of deformation exceeds the tag spacing. This has to be taken into account when choosing parameters such as the tag spacing, the tag delay, and the time scale of the motion of interest.

### 2.3 In vivo datasets

To test the algorithm on in vivo data, two typical shearing deformations were scanned in one healthy, consenting, 26 year old volunteer; deformation of the calf muscles and eye movement. Both the calf muscle contraction and eye movement were voluntary. All scans were acquired using a Philips 3T Intera scanner (Philips Healthcare, Eindhoven, the Netherlands) with a Sense XL 16 channel torso coil for acquisition of the calf muscles and an 8 channel head coil for the eye muscles. Prior to the tagged imaging scans, T1 (leg) or T2 (brain) weighted turbo spin echoes (TSE) were obtained for anatomical reference.

#### 2.3.1 Tagged imaging of the calf muscles

Scan settings of T1 weighted scans: Coronal: voxel size 0.9x9x0.9 mm in plane, slice thickness 4 mm, matrix size: 444x334, TR/TE = 636/16 ms, number of signal averages (NSA) = 2; transverse: voxel size 0.9x 0.9 mm in plane, slice thickness 4 mm, matrix size: 332x334, TR/ TE = 636/16 ms, NSA = 2.

Three tagged series of 60 acquired 3D volumes or dynamic volumes were then acquired with a tag spacing of 8, 12 and 16 mm, all with a tag delay of 150 ms. Acquisition sequence parameters: Turbo Field Echo (TFE), voxel size 3x3 mm in plane, slice thickness 5 mm, matrix size 132x132, TR/TE = 1.96/0.93 ms, FOV = 400x400x30 mm, six slices, flip angle = 8°, SENSE factor 4 in RL direction. The readout time of this acquisition was 98 ms.. The pixel bandwidth was set at 3,788 Hz with a corresponding water fat shift of 0.12 pixels. During the tagged acquisitions, the volunteer was asked to extend and flex the feet in a constant frequency of approximately 1 Hz.

#### 2.3.2 Tagged imaging of eye movement

Scan settings of T2 weighted scans (transverse and sagittal): voxel size 0.45x0.45 mm in plane, slice thickness 3 mm, matrix size: 448x448, TR/TE = 2000/81 ms, NSA = 1. Two tagged series of 60 acquired 3D volumes or dynamic volumes were then acquired, one transverse and one sagittal, all with a tag spacing of 6 mm and a tag delay of 150 ms. Acquisition sequence parameters: TFE, voxel size 1x1.5 mm in plane (1 mm perpendicular to the tag direction and 1.5 parallel), slice thickness 3 mm, matrix size 200x133, TR/TE = 3.3/2 ms, FOV = 200x200x27 mm, nine slices, flip angle = 8°, SENSE factor 3.5 in AP direction. The readout time of this acquisition was 120 ms.. The pixel bandwidth was set at 1,250 Hz with a corresponding water fat shift of 0.348 pixels. During the tagged acquisitions, the volunteer was asked to move his eyes perpendicular to the tag line direction (i.e. from left to right during the transverse acquisition and from up to down during the sagittal acquisition) in a constant frequency of approximately 1 Hz. In all tagged acquisitions the profile order was set to low–high with the readout direction perpendicular to the direction of the tag planes, ensuring a fast and simultaneous readout of the centre and two first harmonic peaks. To obtain SNR, one set was acquired without implementation of a tag pattern. SNR was calculated as the ratio of the signal mean and standard deviation in a

20x20 voxel ROI centered in the FOV over 60 dynamic volumes.

### 2.3.3 Data analysis

The simulated series of shearing deformation in a numerical phantom were segmented using the scale space based algorithm. The segmentation error was calculated as the distance between the segmented and the actual tag points. This distance was averaged over all voxels in the Shepp-Logan phantom and normalized in units of the tag spacing; the success rate S was defined as one minus the segmentation error, *e*:

$$S = 1 - e = \frac{\sum S.L.Phantom |y_i(L_S) - y_i(L_A)}{h \cdot N}$$

(6)

where $y_i(L_S)$ and $y_i(L_A)$ represent the *y* positions of the segmented and actual tags respectively. *N* is the number of voxels in the Shepp-Logan Phantom. In this definition, a success rate of *S* = 1 corresponds to all lines segmented correctly, *S* = 0 corresponds to all segmented lines being off one full tag spacing of the actual lines.

All acquired in vivo series of the calf muscle and eye were segmented with the scale space based algorithm. In the calf muscle data, three dynamic volumes per series were selected containing increasing states of deformation allowing evaluation of the segmentation algorithm at these states.

For each eye movement series, two dynamic volumes with minimal and maximal motion of the eye were selected.

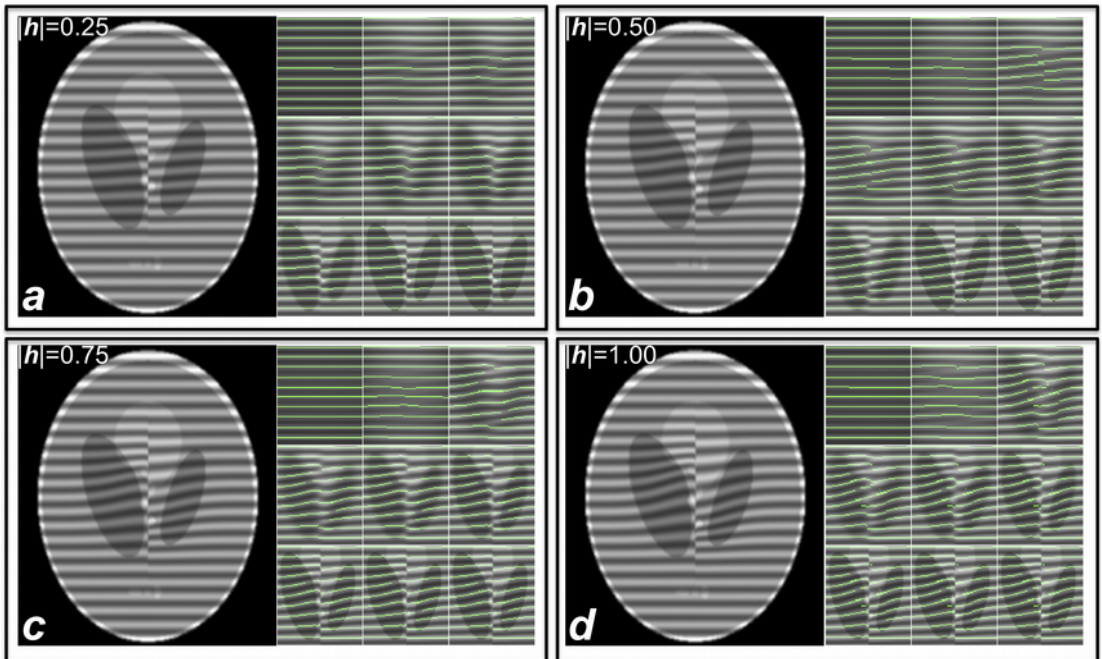

*Figure 3:* Scale space segmentation in the Shepp-Logan phantom at various shift levels **h**. *a-d Shows the segmentation algorithms in several dimensions of scale space for shift |**h**| = 0.25, 0.50, 0.75 and 1.00 d (d= the tag spacing)*

# 3 Results

## 3.1  Numerical phantom dataset

Figure 3 shows excerpts of the scale space based segmentation process at several dimensions of scale space for increasing shifts **h** for the highest SNR level (SNR = ∞). The segmentation can be visibly appreciated as successful in Figure 3a, b. In Figure 3c, small errors occur around the centre of the image. In Figure 3d, where the shift |**h**| = $d$ these errors have increased. In Figure 3c and d the net separation in the centre of the image equals 1.5 and 2 times the tag spacing because of the shearing effect. The deformation thus exceeds the tag spacing in the origin of Figure 3c, and in a larger area around the origin in Figure 3d. As a consequence, spurious edges occur at various scales causing the segmentation process to fail where separate tags blend. In Figure 3a, blending does not occur at ay scale and the tagged voxels are segmented correctly, detaching along the shearing interface in the centre of the image. In Figure 3b, a small amount of blending is present at intermediate scales. The segmentation is however still successful, which can be attributed to the high SNR of the image.

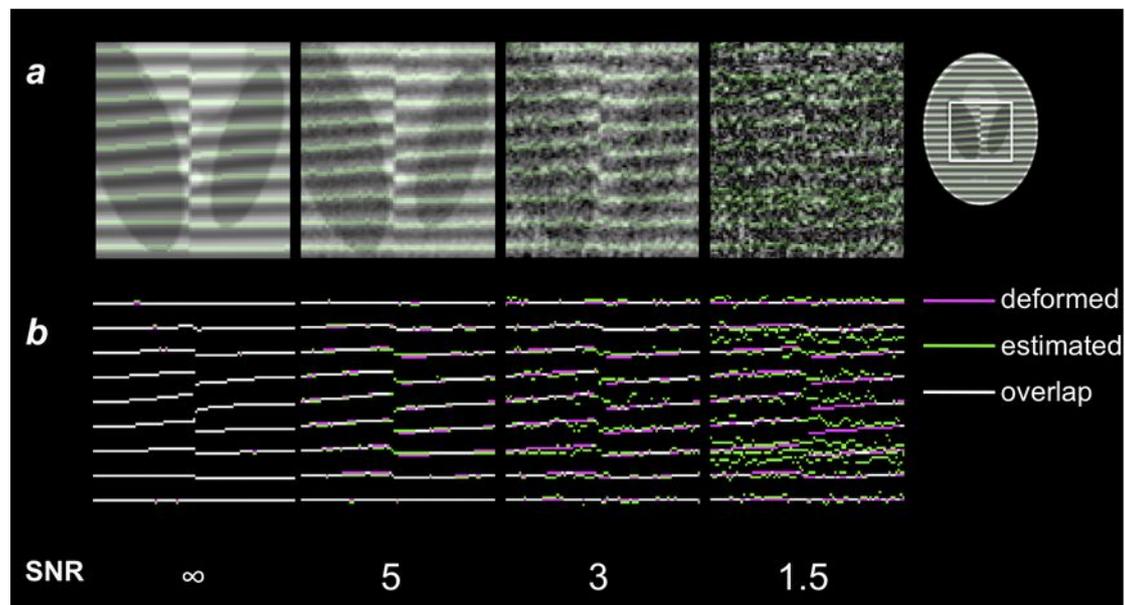

**Figure 4:** *Segmentation results from the simulation.* **a** *Shows the deformed Shepp-Logan phantom at a shift of |**h**| = 0.25d for different SNR levels. The segmented tags are overlaid in green (please refer to the online version for color figures).* **b** *Shows the correspondence of the segmented results with the original imposed deformation. The segmentation results are indicated in green, the correct lines are indicated in magenta, white means magenta and green coincide and the segmentation is thus correct.*

Figure 4a shows the phantom dataset at a shift of |**h**| = 0.25 $d$ at several noise levels. At SNR = ∞ (i.e. no added noise) the algorithm segments the shearing tagged voxels with negligible error. Note that even around the discontinuous shearing edge the tagging pattern is successfully segmented. The error in the segmented tags increases with decreasing SNR. At SNR = 1.5 the algorithm clearly fails in segmenting the tagging pattern. Figure 5 shows the success rate as a function of the displacement h and SNR. At SNR = ∞, the success rate approaches 1 for a shift of up 30 % of the tag spacing. Note: this equals a net separation of 60 % of the tag spacing in the middle

of the phantom because of the shearing effect. The success rate suddenly drops after 30 % because the deformation starts to approach the tag spacing, leading to the creation of spurious edges in the scale space algorithm (see section 2.2), in turn resulting in segmentation errors. The same drop in success can be appreciated for all SNR levels. For SNR = 10, the success rates remain above 0.97 for shifts up to 30 %.

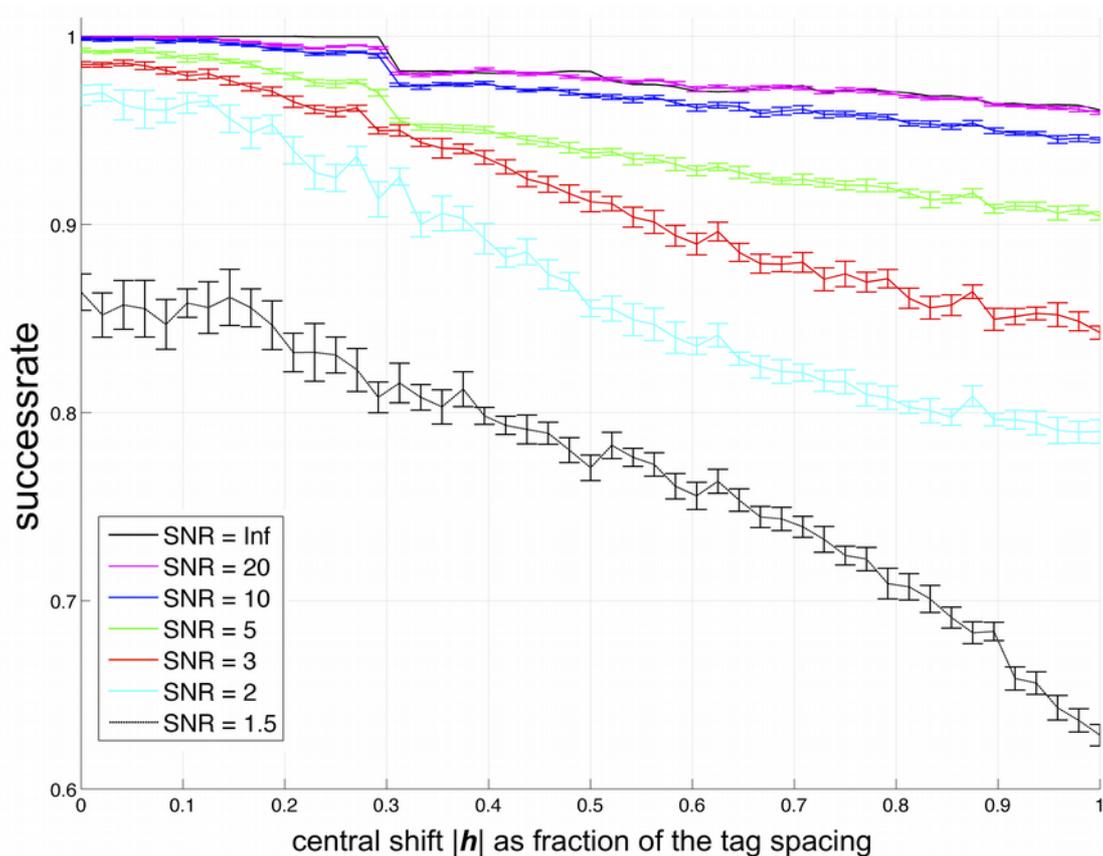

*Figure 5:* Success rate of the segmentation algorithm on simulated shearing as a function of the imposed shift **h** at several SNR levels, with error bars denoting the standard deviation over repeated measurements

### 3.2  In vivo dataset

The images were acquired successfully. Figure 6 shows tagged dynamic images in the calf muscle from the in vivo acquisitions in one healthy volunteer. Coronal and transverse T1 weighted images are added for anatomical reference (Figure 6a). The SNR was calculated over the two regions indicated in the first image of Figure 6b at SNR = 18.7.

Overall, tagged acquisition and subsequent segmentation was most successful in the middle part of the FOV where B0 artifacts were minimal. In Figure 6b, three states of deformation are presented for the tagged series with tag spacing 8, 12, and 16 mm. All images are displayed with and without segmented tags (first three and second three columns). For each image, two areas at the interface of the soleus muscle and gastrocnemius muscle are enlarged.

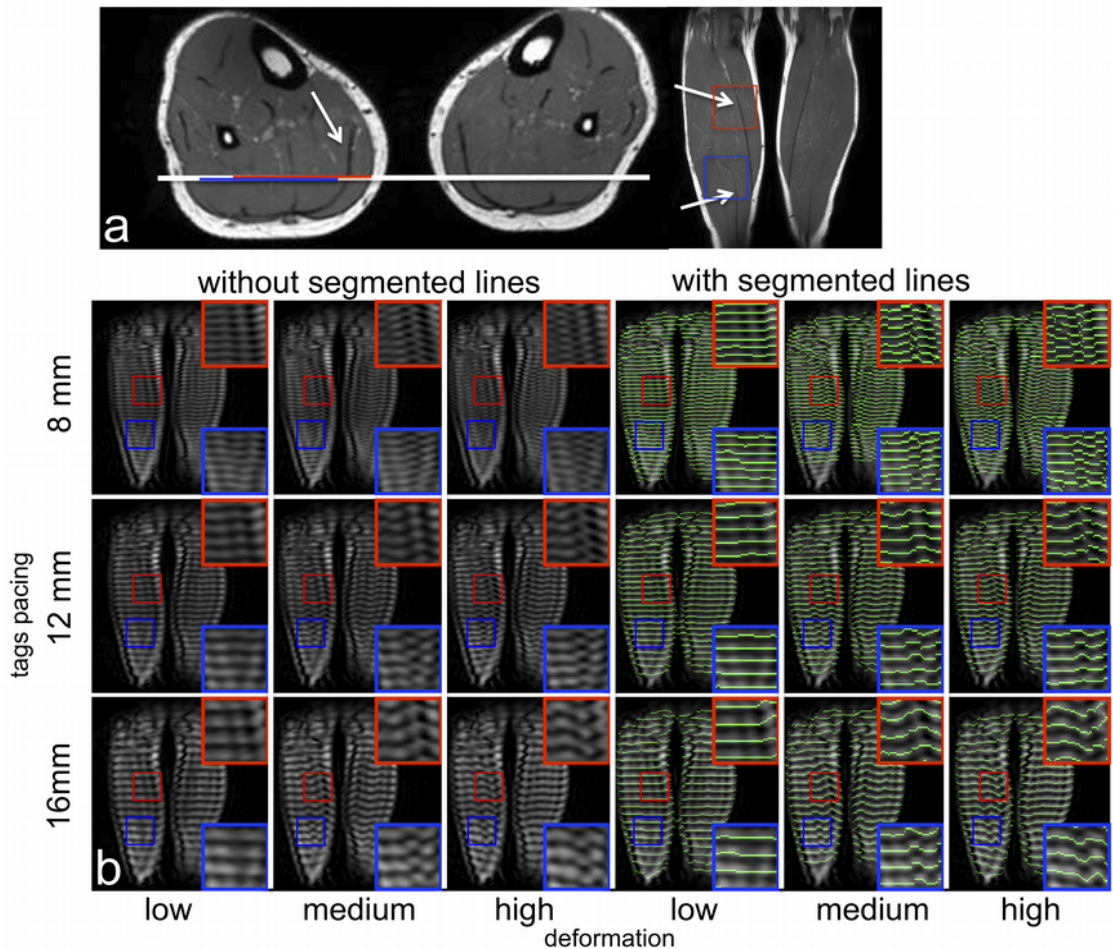

*Figure 6:* Excerpts of the three in vivo tagged series of the deforming calf muscle in one healthy volunteer (**b**). Interface areas of the soleus muscle and gastrocnemius muscle are enlarged in the side frames of each image. Coronal and axial T1 weighted images are displayed in **a** for anatomical reference. Soleus and gastrocnemius muscle interfaces are indicated with white arrows. The two enlarged areas of muscle interfaces are indicated with red and blue

In the 8 mm series, segmentation appears valid in these areas for the first and second state of deformation, but fails at the third state. Here, the deformation clearly exceeds the tag spacing. For the 12 mm tag spacing series segmentation appears valid for all three states of deformation. The segmented tagged voxels detach at the soleus-gastrocnemius interface, where shearing can be visibly appreciated. The 16 mm series also appear correctly segmented for all states of deformation. Because of the larger tag spacing, any possible shearing motion has not caused the tagged voxels to detach.

It can be noticed that due to the large bandwidths no displacement is present in the tagged voxels from the fat to the muscle regions. Chemical shift artifacts of the second kind can be observed at all water fat transitions in the image, since in our sequence an almost out-of-phase TE was used.

Figure 7a and b show mid-slice images of the T2 weighted transverse and sagittal acquisition for anatomical reference of the tagged acquisitions. Figure 7c show tagged acquisitions of the transverse series with minimal and maximal motion (left and right) displayed with and without overlay of the segmented lines (up and down). Analogous to Figure 7c, d shows two tagged acquisitions with minimal and maximal motion of the sagittal series. Average image SNR was calculated in the tagged acquisitions over the area indicated in the anatomical reference images (Figure 7a, b) to be 15.4 for the transverse series and 15.1 for the sagittal series. Because of the low resolution of the tagged acquisitions, it is difficult to distinguish the different muscles controlling the eye movement and the eyeball itself. From the transverse tagged images with minimal motion in the left side of Figure 7c it can be seen that tag lines detach in the areas containing eye muscle/eyeball interface. On the right side of Figure 7c, maximal movement of the eye causes more breaking of taglines crossing the eye muscle/eyeball interface. On the left side of Figure 7d, the sagittal tagged images with minimal motion show virtually no breaking of the tag lines. In the sagittal tagged images with maximal motion on the right side of Figure 7d, the taglines shear and break near the eye muscle/eyeball interface. The scale space based algorithm segments the tag lines correctly in all images. In this acquisition, an in-phase TE was used, which can be appreciated from the absence of chemical shift artifacts of the second kind (see also above).

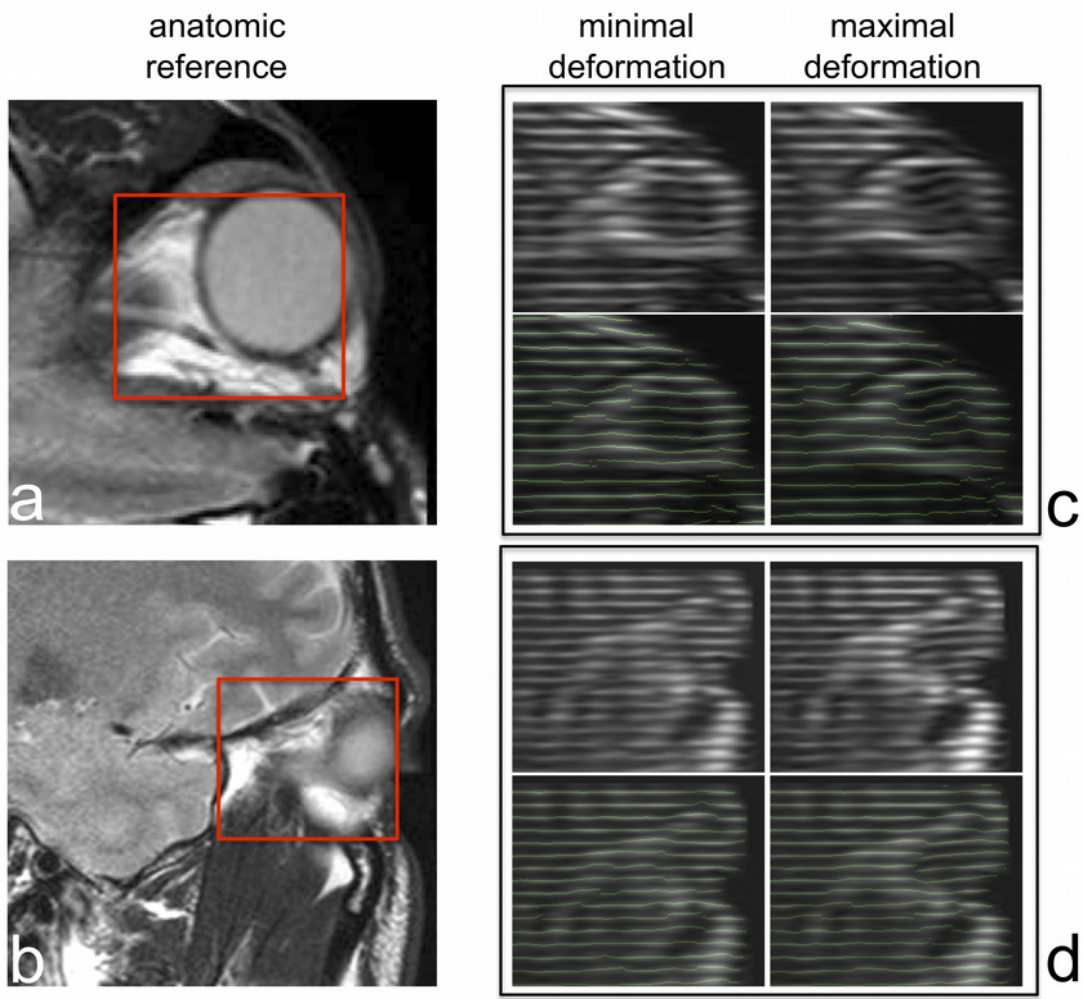

*Figure 7:* Tranverse and sagittal tagged acquisitions of eye movement. The healthy volunteer was asked to move his eyes from left to right during the transverse acquisition (***a,c***) and from up to down during the sagittal (i.e. parallel to the optic nerve) acquisition (***b,d***). ***a,b*** show anatomic reference for the tagged acquisition in ***c,d*** respectively. From the two sets of 60 dynamic volumes, images with maximal and minimal motion caused by the constant eye movement were selected. Motion is minimal on the left side of ***c*** and ***d*** and maximal on the right side.

# 4 Discussion

Single shot tagged MRI of complex deformation can be segmented automatically with high precision using the scale space based algorithm, provided that the deformation is smaller than the tag spacing. The amount of deformation can be adjusted by changing either the tag spacing or the tag delay.

With the simulated shearing edge deformation, the algorithm's success rate was methodically evaluated by varying the amount of shearing deformation and SNR level. This showed a success rate of 0.97 and higher for deformation that remains within the tag spacing at SNR = 10 and higher. The algorithm thus functions well at moderate SNR levels, which are to be expected in single shot tagged MRI. The *in vivo* data qualitatively confirmed these results and showed that the tag spacing can be varied and optimized for maximum deformation with successful segmentation of the deformed taglines. Although this algorithm was developed for single shot tagged acquisitions of non-periodic motion, it could also be applied to triggered, tagged cardiac data.

Alternative approaches to be considered include energy minimization algorithms such as snakes [17–20]. These can work well on segmentation of tagged data but have trouble with shearing edges. Snake-like algorithms also require a set of initial contours or template patterns from which the energy minimization procedure can be started. Advanced analysis methods for cardiac tagged data such as HARP [21, 22] exploit the advantages of triggered imaging of periodic motion and employ band-pass filtering in the *k* space domain. As a consequence, accurate decomposition of the tag peaks and the central or DC peak in k space containing the anatomical image (C-SPAMM [13, 14]) ensures good results. The scale space based algorithm can segment breaking tags at relatively low SNR levels and does not require the initial locations of the tags. It does not employ filtering in the *k* space domain but rather suppresses the underlying anatomical image through blurring in the spatial domain. Image manipulation of the tag pattern in the *k* space domain without the option of decomposing the central peak from the higher order tagging peaks can be prone to errors. This is especially true when the low spatial resolution enforces a large tag spacing, in turn reducing the distance between the DC peak and the first order tagging peaks. Its robust and automatic nature offers great potential for processing of large datasets and enables effective post-processing of complex tagged deformation.

In this study, 1-1 tag patterns were used in both the numerical simulation and the in vivo experiments. The scale space based algorithm exploits the tag pattern characteristic of having maximum intensity on the location of each tag line and therefore works best on these first order tag patterns. For successful segmentation, the deformation in the single shot tagged data must not exceed the tag spacing, as this will cause spurious edges in several scale space dimensions, which in turn will result in segmentation errors. Although the tag spacing can be chosen such that excess is not likely to occur, depending on the complexity and predictability of the motion of interest, this excess cannot be ruled out.
Furthermore, through-plane motion and field inhomogeneities can also cause blending of tags and thus segmentation errors. The algorithm presented here was developed for 1D tag patterns, and should be expanded in future research to 2D and 3D tag patterns.

## 5 Conclusion

The scale space based algorithm for automatic segmentation of tagged images presented here is a robust algorithm that enables the application of tagged MR to complex (shearing) deformation. It can process large datasets with relatively low SNR. The algorithm is particularly suited for non-triggered, single shot tagging, since it does not require initial contours or filtering of harmonic peaks in $k$ space.